\documentstyle[12pt]{article}

\input{tcilatex}
\begin{document}

{\footnotesize {hep-th/9607112 {\hfill } USC-96/HEP-B4}} \bigskip

\begin{center}
{\bf S-THEORY}

\baselineskip=22pt
\end{center}

\centerline{\footnotesize ITZHAK BARS} \baselineskip=13pt

\centerline{\footnotesize {\it Department of Physics and Astronomy,
University of Southern California}} \baselineskip=12pt \centerline{%
{\footnotesize {\it Los Angeles, CA 90089-0484, USA}}}

\centerline{\footnotesize E-mail: bars@physics.usc.edu}

\vspace*{0.9cm} {\ \centering{%
\begin{minipage}{12.2truecm}\footnotesize\baselineskip=12pt\noindent
\centerline{\footnotesize ABSTRACT}\vspace*{0.3cm}
\parindent=0pt \

The representation theory of the maximally extended superalgebra with 32 fermionic 
and 528 bosonic generators is developed in order to investigate non-perturbative 
properties of the democratic secret theory behind strings and other p-branes. 
The presence of Lorentz non-singlet central extensions is emphasized, their 
role for understanding up to 13 hidden dimensions and their physical
interpretation as boundaries of p-branes is elucidated. The criteria for a new
larger set of BPS-like non-perturbative states is given and the methods of
investigation are illustrated with several explicit examples.

\end{minipage}}}


\baselineskip=15pt \setcounter{footnote}{0} \renewcommand{\thefootnote}{%
\alph{footnote}}


\section{Introduction}

String and p-brane dualities have led to the notion that there is a {\bf m}%
ysterious, {\bf f}ascinating, {\bf s}ecret {\bf s}upersymmetric theory,
which may be called M-theory, F-theory, S-theory, $\cdots $. It includes all
possible closed and open (i.e. with boundaries) p-branes, such as {\bf s}%
trings, {\bf m}embranes, {\bf f}ivebranes, $\cdots ,$ interacting with each
other in various flat, curved or compactified spacetimes. It has been
proposed that this theory, which includes up to 12 (or 13, see below) hidden
dimensions, may be the {\bf m}other, {\bf f}ather or {\bf s}ire of all
physical theories. Some of the M- and F- aspects have been discussed by
other authors \cite{witten}\cite{vafaf}\cite{jhs}. D-branes \cite{pol}
provide some handle on the theory.

In this paper I will concentrate on some of the S- properties of the {\bf s}%
ecret theory, and will emphasize an algebraic approach \cite{ibbeyond} based
on a superalgebra with 32 fermionic and 528 bosonic generators \cite
{townsend}\cite{ibbeyond}. The collection of all bosonic operators form a 32$%
\times 32$ symmetric matrix $S$ given by the anticommutator of the
supercharges $\left\{ Q,Q\right\} \sim S$. The structure and symmetries of $%
S $ are related to $p$-branes, dualities and hidden dimensions. Global
properties, certain states and certain non-perturbative properties of the
underlying secret theory may be studied by analyzing the representations of
this superalgebra. This line of investigation will be called S-theory.

In this paper I will outline the elements of S-theory. One of the points to
be emphasized is the presence of Lorentz non-singlet central extensions in $%
S,$ which so far received little attention$.$ In the usual treatment of
non-perturbative properties of M- or F-theory the Lorentz singlet central
extensions play a major role (e.g. BPS-like states or black holes). Here I
will argue that the Lorentz non-singlets also play a major role and that
their presence is required in order to see the full extent of hidden
dimensions and duality symmetries. In section-2 I will give a physical
interpreation of these central extensions in terms of boundary variables for 
$p$-branes. In section-3 the symmetry structure of the generalized algebra
and the connection to hidden dimensions (up to a total of 13 dimensions)
will be explained. In section-4 the representations of the superalgebra will
be constructed and, as a special case, the criteria for obtaining the
quantum numbers of the generalized BPS-type states will be given. As an
application of S-theory I will show that there are new non-perturbative BPS
states that carry quantum numbers (eigenvalues of central extensions) that
are Lorentz non-singlets, and I will construct some explicit examples. Since
these quantum numbers have an interpretation in terms of $p$-brane
boundaries these BPS states are related to open $p$-branes. When they are
included in the spectrum along with more familiar D-brane BPS states they
form {\it larger multiplet structures of symmetries} that exhibit additional
hidden dimensions as well as dualities.

Results such as the ones described, which follow only from the properties of
the superalgebra, are assumed to be valid non-perturbatively in the full
theory. Therefore, they must be useful handles for constructing and
analysing the fundamental underlying theory.

\section{Lorentz non-singlet central extensions}

In the generalized superalgebra $\left\{ Q_\alpha ^a,Q_\beta ^b\right\}
=S_{\alpha \beta }^{ab},$ where $S_{\alpha \beta }^{ab}=\delta ^{ab}$ $%
\gamma _{\alpha \beta }^\mu \,P_\mu +$ ``central extensions'' , there are
generally Lorentz non-singlet central extensions $Z_{\mu _1\cdots \mu
_p}^{ab}$ with $p$ Lorentz indices. The structure and properties of these
additional operators are explained more fully in the following sections. As
explained in \cite{ibbeyond} a nonzero $Z_{\mu _1\cdots \mu _p}^{ab}$ which
is not a Lorentz singlet does not violate the no-go theorem of \cite{haag}
as long as there are extended objects. The Lorentz singlets $Z^{ab},$ with $%
p=0,$ are well known to represent the quantum numbers of black holes in M-
theory. In this section I will clarify the physical interpretation of $%
Z_{\mu _1\cdots \mu _p}^{ab}$ for $p\geq 1$ as related to $p$-branes.

For simplicity I will assume that all $p$-branes propagate in flat
backgrounds (such as flat spacetime in direct product with tori or their
orbifolds in compactified dimensions). Under this assumption all 528
components of $S$ commute with each other as justified below. Similar
considerations in curved backgrounds would yield a more complicated algebra
that is harder to analyse (for example momenta do not commute in curved
backgrounds).

Just like momentum, all possible values of these central extensions must be
included in the representation of the superalgebra in order to take into
account all possible states in the representation consistent with the
Lorentz group. But physical considerations would determine if they are
spacelike, lightlike or timelike, and hence there are classes of
representations, just like the Poincar\'{e} group.

These central extensions are closely associated with the boundaries of $p$%
-branes as well as the topology of the background geometry in which they
propagate. In a flat $d$-dimensional spacetime, as assumed in this paper,
the Lorentz non-singlets are present only if the $p$-brane has boundaries.
This is seen as follows: In the low energy effective theory the gauge
potential with $p+1$ antisymmetric Lorentz indices has a source term in its
equation of motion (the notation is explained in \cite{ibbeyond}) 
\begin{equation}
\partial _{\lambda \,\,}\partial ^{[\lambda }A_{ab}^{\mu _0\mu _1\cdots \mu
_p]}\left( x\right) =J_{ab}^{\mu _0\mu _1\cdots \mu _p}\left( x\right) .
\label{source}
\end{equation}
The current is non-vanishing when constructed from a $p$-brane $X_\mu (\tau
,\sigma _1,\cdots ,\sigma _p)$ and its super-partners 
\begin{eqnarray}
J_{\mu _0\mu _1\cdots \mu _p}^{ab}\left( x\right) &=&\int d\tau d\sigma
_{1\cdots }d\sigma _p\sum_iz_i^{ab}\,\delta ^d\left( x-X^{\left( i\right)
}(\tau ,\sigma _{1,}\cdots \sigma _p)\right) \,\,  \label{current} \\
&&\times \partial _\tau X_{[\mu _0}^{\left( i\right) }\cdots \,\partial
_{\sigma _p}X_{\mu _p]}^{\left( i\right) }(\tau ,\sigma _{1,}\cdots \sigma
_p)\,\,+\cdots  \nonumber
\end{eqnarray}
where the index $i$ is a label for many $p$-branes and $z_i^{ab}$ is their
coupling to the pair of supercharges labelled by $a,b$. The Lorentz
non-singlet central extension is the integral of the current over a
spacelike ``slice'' in spacetime 
\begin{equation}
Z_{\mu _1\cdots \mu _p}^{ab}=\int d^{d-1}\Sigma ^{\mu _0}\,\,J_{\mu _0\mu
_1\cdots \mu _p}^{ab}\left( x\right) .  \label{charge}
\end{equation}
where, e.g. one may use a non-covariant notation by choosing $\mu _0=0,$ $%
d^{d-1}\Sigma ^0=d^{d-1}x$, i.e. the volume of space at fixed time. This is
one expression for $Z_{\mu _1\cdots \mu _p}^{ab}.$ Another expression is
obtained by substituting the left side of (\ref{source}) in (\ref{charge}).
Then the integrand is a total divergence and therefore it can be expressed
as a surface integral involving the asymptotic values of $A_{ab}^{0\mu
_1\cdots \mu _p}$ at infinity of physical space ($r\rightarrow \infty )$.
Therefore in any classical solution of the effective low energy theory, a
non-trivial asymptotic behavior $A_{ab}^{0\mu _1\cdots \mu _p}\sim
Z_{ab}^{\mu _1\cdots \mu _p}\,/\,r^{d-2}$ would have an interpretation in
terms of $p$-branes through (\ref{current}).

Now, what property of the $p$-brane is represented by $Z_{\mu _1\cdots \mu
_p}^{ab}?$ As an example, let us perform the integral in (\ref{charge}) for $%
p=1,$ i.e. for a string. Ignoring the $\cdots $ due to fermionic string
variable (due to supersymmetry), the result is\footnote{%
It is useful, but not necessary to choose the timelike gauge $X^0\left( \tau
,\sigma \right) =\tau ,$ use $\tau =x^0$ because of the delta function, and
take the spacetime ``surface'' to be the usual integral over all space at
constant time.} 
\begin{equation}
Z_\mu ^{ab}=\sum_iz_i^{ab}\int d\sigma \,\,\partial _\sigma X_\mu ^{\left(
i\right) }\left( \tau ,\sigma \right) .
\end{equation}
If the remaining integral over $\sigma $ is for closed strings propagating
in flat spacetime, then the closed string condition $X_\mu ^{\left( i\right)
}\left( \tau ,0\right) =X_\mu ^{\left( i\right) }\left( \tau ,2\pi \right) $
gives ($Z_\mu ^{ab})_{closed}=0\,\,\,$\footnote{%
A similar expression in compactified dimensions is non-zero because of the
closed cycles in a non-trivial topology. Similarly, if spacetime is curved
instead of flat, there could be non-trivial contributions for closed
strings, if the topology is non-trivial.}. But if there are open strings,
then the the result depends only on the end points 
\begin{equation}
\left( Z_\mu ^{ab}\right) _{open}=\sum_iz_i^{ab}\left( X_\mu ^{\left(
i\right) }\left( \tau ,\pi \right) -X_\mu ^{\left( i\right) }\left( \tau
,0\right) \right) .
\end{equation}
Note that this is a Lorentz vector, and (unless identically zero) it is a
continuous spacelike or light-like variable, but is not timelike. It is
obviously translationally invariant and hence it commutes with the momentum
operator, consistent with our general assumption that all central extensions
commute in flat spacetime.

Similarly, for the general $p$-brane in flat spacetime the central
extensions $Z_{\mu _1\cdots \mu _p}^{ab}$ can be shown to be related to
boundary variables that commute with each other as well as with the momentum
operators. Hence all these operators are simultaneously diagonalizable and
their continuous eigenvalues must label the physical states at an equal
footing with the eigenvalues of momentum, since they are not distinguishable
under the symmetries of the superalgebra.

\section{The extended superalgebra}

The maximum number of supercharges in a physical theory is 32. This
constraint comes from 4-dimensions, which admits at the most 8 supercharges,
each with four real components, since there can be no supermultiplet of
massless particles with spins higher than 2. In an arbitrary number of
dimensions, we label the 32 supercharges as 
\begin{equation}
Q_\alpha ^a=\left\{ 
\begin{array}{l}
\alpha \,\,\text{=spinor in }d\text{ dims}. \\ 
a=1,2,\cdots N;\quad \text{spinor in }c+2\text{ dims.}
\end{array}
\right.
\end{equation}
where $N=1$, when $d$=11; $N=2,$ when $d$=10;$\cdots $;$\,\,N=8$ when $d$=4.
Here $d$ is the dimension of spacetime. Let's define $c$ as the number of
compactified string dimensions such that $d+c=10.$ It was argued that there
are two extra hidden dimensions, one spacelike and one timelike, and that $N$
corresponds to the dimension of the spinor in $c+2$ dimensions \cite
{ibbeyond}. Furthermore $N$ also corresponds to a dimension of an
irreducible representation of the group $K,$ which is the maximal compact
subgroup of $U$-duality.

The extended superalgebra has the form 
\begin{equation}
\begin{array}{c}
\left\{ Q_\alpha ^a,Q_\beta ^b\right\} =\left( S\right) _{\alpha \beta }^{ab}
\\ 
\left( S\right) _{\alpha \beta }^{ab}=\delta ^{ab}\gamma _{\alpha \beta
}^\mu \,\,P_\mu +\sum_{p=0,1,,2,\cdots }\gamma _{\alpha \beta }^{\mu
_1\cdots \mu _p}\,\,\,Z_{\mu _1\cdots \mu _p}^{ab}.
\end{array}
\label{ab}
\end{equation}
where the permutation symmetry of ($\alpha \beta )$ must be the same as $%
\left( ab\right) $ in each term on the right hand side. The structure of $%
\left( S\right) _{\alpha \beta }^{ab}$ is of central interest in this paper.
From it we will learn about the symmetries of the underlying theory (many of
them hidden from the point of view of conventional string theories) as well
as about the representation space that is related to the physical states of
the theory.

For $d\geq 10$ there are two types of superalgebras A,B that are outlined
below. From the construction of gamma matrices given in the appendix one can
see that the type A and type B superalgebras originate with different forms $%
S_A,S_B$ embedded in a higher 64$\times 64$ spinor space in 13-dimensions.
For $d\leq 9$ there is a one to one correspondance between type A,B
superalgebras, which are then related by T-duality. As we will see below
certain hidden symmetries in the form (\ref{ab}) become manifest in the
covariant forms $S_A,S_B$ given below in higher dimensions. Hence the
T,U-duality between A and B versions may find its roots in 13D.

The Lorentz generators for $SO(d-1,1)$ do not commute with the supercharges
or with the bosonic generators (except for $p=0$ case), since these are
spinors or antisymmetric $p$-tensors under Lorentz transformations
respectively. In addition to the Lorentz transformations that act as
isometries, there are other isometries that act like (i) Lorentz
transformations in $c+2$ compact plus hidden dimensions $SO(c+1,1)$ and (ii)
Discrete $U-$duality transformations that induce continuous transformations
through the maximal compact subgroup $K$ (see \cite{ibbeyond}). The
supercharges, bosonic generators and the physical states of the theory are
classified by the isometries $SO(d-1,1)\otimes $ $SO(c+1,1)$ or by $%
SO(d-1,1)\otimes K$. Displaying one of these classifications may hide the
other one. The isometries and their intersecting subgroup structure was
given in \cite{ibbeyond} as follows 
\begin{equation}
\begin{array}{l}
\left. 
\begin{array}{c}
\left. 
\begin{array}{c}
{\small c\,\,}\text{{\small compact\thinspace \thinspace +}} \\ 
{\small 2\,}\text{{\small hidden dims}}
\end{array}
\right. \\ 
SO(c+1,1) \\ 
\downarrow
\end{array}
\right. \quad \rightarrow \quad \,\,\,\,\,\,\left. 
\begin{array}{c}
\text{{\small two \thinspace intersecting}} \\ 
\text{{\small classifications}} \\ 
\text{{\small of multiplets\thinspace \thinspace of}} \\ 
\text{{\small generators \&\thinspace \thinspace states}}
\end{array}
\right. \\ 
\left. 
\begin{array}{l}
\left. 
\begin{array}{l}
SO(c+1)_{1\,\,hidden\,\,\dim .} \\ 
{\small SO(c)\,\,is\,\,common\,\,subgroup} \\ 
SO(c)_L\otimes SO(c)_R
\end{array}
\right\} \rightarrow \left. 
\begin{array}{c}
\uparrow \\ 
K \\ 
\text{{\small maximal compact in }}U
\end{array}
\right. \\ 
\begin{array}{l}
\,\,\,\,\,\,\,\,\,\,\,\,\,\,\,\,\,\,\,\,\,\,\,\,\,\,\,\,\uparrow \\ 
T\,-\,{\small duality}\,\,SO(c,c)
\end{array}
\end{array}
\right\} \rightarrow \stackunder{duality}{U}
\end{array}
\label{subgroups}
\end{equation}

In this paper I will assume a flat background and hence all bosonic
generators commute among themselves and with the supercharges as explained
in section-2. In this case it is easy to find all the representations of the
superalgebra and analyse the physical states as discussed in section-4.

\subsection{Type-A}

In 11D there are only 3 terms \cite{townsend} 
\begin{equation}
\begin{array}{c}
\left\{ Q_\alpha ,Q_\beta \right\} =\left( S_A\right) _{\alpha \beta } \\ 
\left( S_A\right) _{\alpha \beta }=\left( C\gamma ^m\right) _{\alpha \beta
}\,\,P_m+\left( C\gamma ^{m_1m_2}\right) _{\alpha \beta
}\,\,\,Z_{m_1m_2}+\left( C\gamma ^{m_1\cdots m_5}\right) _{\alpha \beta
}\,\,\,X_{m_1\cdots m_5}.
\end{array}
\label{type2}
\end{equation}
where $m=0,1,2,\cdots 10,$ and the gamma matrices are 32$\times 32$ (for
more details see the appendix). When reduced to lower dimensions this 32$%
\times 32$ matrix $\left( S_A\right) _{\alpha \beta }$ takes the form of (%
\ref{ab}). As is well known, in 10D $\left( S_A\right) _{\alpha \beta }$ is
distiguishable from the type-B $\left( S_B\right) _{\alpha \beta }$ given
below, but in 9 dimensions or less the reduced $\left( S_A\right) _{\alpha
\beta }$ and reduced $\left( S_B\right) _{\alpha \beta }$ have similar
content that is related by T-duality.

It is also possible to consider 12 dimensions with signature (10,2) \cite
{duff} since the Weyl spinor is real and 32-dimensional. Then the extended
superalgebra can be written covariantly in 12D \cite{ibbeyond}

\begin{equation}
\begin{array}{c}
\left\{ Q_\alpha ,Q_\beta \right\} =\left( S_A\right) _{\alpha \beta } \\ 
\left( S_A\right) _{\alpha \beta }=\left( C\gamma ^{M_1M_2}\right) _{\alpha
\beta }\,\,\,Z_{M_1M_2}+\left( C\gamma ^{M_1\cdots M_6}\right) _{\alpha
\beta }\,\,\,Z_{M_1\cdots M_6}^{+}
\end{array}
\label{type2a}
\end{equation}
where $M=0^{\prime },m,$ and $m=0,1,2,\cdots 10,$ with two time-like
dimensions denoted by $M=0^{\prime },$ $0.$ The relation between 11 and 12
dimensional gamma matrices $\gamma _{\alpha \beta }^m,\gamma _{\alpha \beta
}^M,$ $\Gamma _{\alpha \beta }^M$ is given in the appendix. Similarly, the
bosonic generators are related by 
\begin{equation}
\begin{array}{l}
Z_{M_1M_2}\rightarrow P_m\oplus Z_{m_1m_2}\quad 66=11+55 \\ 
Z_{M_1\cdots M_6}^{+}\rightarrow X_{m_1\cdots m_5}\quad \quad \quad 462=462.
\end{array}
\end{equation}
The six-index tensor is self dual in 12D.

Note that there is no 12D translation operator $P_M$ in the (10,2) version (%
\ref{type2a}). Therefore the extension of the theory from (10,1) to (10,2)
is {\it not the naive extension that would have implied two time coordinates}
, since the corresponding canonical conjugate momenta are not present in the
theory. There is only one time translation operator $P_0$, hence there is
only one time coordinate that can be recognized after the reduction to 11 or
lower dimensions. Nevertheless there is an obvious SO(10,2) covariance in
the form (\ref{type2a}). Therefore there is a hidden SO(10,2) covariance in
the forms (\ref{type2}) or in (\ref{ab}) for $d\leq 9,$ since they are
equivalent to (\ref{type2a}). We emphasize again that the 12D SO(10,2)
generators $L_{MN}$ do not appear on the right hand side; i.e. $Z_{M_1M_2}$
are not the $L_{MN}.$

\subsection{Type-B}

The type-B superalgebra in $d=10$ can be written as (see appendix) 
\begin{equation}
\begin{array}{c}
\left\{ Q_{\bar{\alpha}}^{\bar{a}},Q_{\bar{\beta}}^{\bar{b}}\right\} =\left(
S_B\right) _{\bar{\alpha}\bar{\beta}}^{\bar{a}\bar{b}} \\ 
\left( S_B\right) _{\bar{\alpha}\bar{\beta}}^{\bar{a}\bar{b}}=\bar{\gamma}_{%
\bar{\alpha}\bar{\beta}}^{\bar{\mu}}\,\,\left( c\bar{\tau}_i\right) ^{\bar{a}%
\bar{b}}P_{\bar{\mu}}^i+\bar{\gamma}_{\bar{\alpha}\bar{\beta}}^{\bar{\mu}_1%
\bar{\mu}_2\bar{\mu}_3}\,\,c^{\bar{a}\bar{b}}\,Y_{\bar{\mu}_1\bar{\mu}_2\bar{%
\mu}_3}+\bar{\gamma}_{\bar{\alpha}\bar{\beta}}^{\bar{\mu}_1\cdots \bar{\mu}%
_5}\,\,\,\left( c\bar{\tau}_i\right) ^{\bar{a}\bar{b}}X_{\bar{\mu}_1\cdots 
\bar{\mu}_5}^i.
\end{array}
\label{type2b}
\end{equation}
where $\bar{\alpha},\bar{\beta}=1,2,\cdots ,16$ and $\bar{a},\bar{b}=1,2$
while $\bar{\mu}=0,1,\cdots ,9$ and $i=0^{\prime },1^{\prime },2^{\prime }.$
The $X_{\mu _1\cdots \mu _5}^i$ are self dual in 10D. Here $\bar{\gamma}_{%
\bar{\alpha}\bar{\beta}}^{\bar{\mu}}\,$are 16$\times 16$ 10D gamma matrices
obtained from the list of 13D gamma matrices in (\ref{gammas}) by omitting
the first two factors in the direct products (i.e. $\gamma _0=1\otimes 1_8,$ 
$\gamma _9=\tau _3\otimes 1_8,$ etc.). The $\bar{\tau}_i^{ab}\equiv \left(
i\tau _2,\tau _3,\tau _1\right) $ are 2$\times 2$ gamma matrices in a hidden
3D Minkowski space, with a charge conjugation matrix $\,c^{ab}=i\tau
_2^{ab}=\varepsilon ^{ab}.$ They are obtained from the 13D gamma matrices in
the appendix by keeping only the first factor in $\left( \Gamma _{0^{\prime
}},\Gamma _{10},\Gamma _A\right) $ which corresponds to $\bar{\tau}_i.$ The
reason for the split of 13D into 10D+3D is the type B chiral projection, as
explained in the appendix. The 13D covariance is lost because of the
projection, but a clear identification of the dimensions labelled by $\bar{%
\mu},i$ survives. This algebra is covariant under SO(1,9)$\otimes SO(1,2)_B.$

Again, as in the type A algebra, there is only one time translation
operator. The usual 10D momentum operator $P_{\bar{\mu}}$ corresponds to the 
$i=0^{\prime }$ component of $P_{\bar{\mu}}^i,$ or equivalently to the trace
part of $\left( c\bar{\tau}_i\right) ^{ab}P_{\bar{\mu}}^i.$ Likewise, the 3D
momentum operator $P^i$ corresponds to the the $\bar{\mu}=0$ component of $%
P_{\bar{\mu}}^i$ or to the trace part of $\bar{\gamma}_{\bar{\alpha}\bar{%
\beta}}^{\bar{\mu}}\,\,P_{\bar{\mu}}^i.$ The time translation operator for
either 10D or 3D is the same one, namely $P_0^{0^{\prime }},$ and it
corresponds to the trace part of the full 32$\times 32$ matrix $\left(
S_B\right) _{\bar{\alpha}\bar{\beta}}^{\bar{a}\bar{b}}.$ This is the same
time translation operator that appears as the trace of $S_A\,$in the type A
superalgebra, given above in the a 12D or 11D covariant forms.

\subsection{More about the duality map A$\Leftrightarrow $B}

Once the theory is compactified to $d=9$ (or fewer dimensions) the two types 
$S_A,S_B$ reduce to two forms that are in one to one correspondance to each
other, but are not identical. Both of these forms display SO(8,1)$\otimes $%
SO(2,1) isometry, one in the form SO(8,1)$\otimes $SO(2,1)$_B$ coming from
SO(9,1)$\otimes $SO(2,1)$_B$ and the other SO(8,1)$\otimes $SO(2,1)$_A$
coming from SO(10,2)$.$ By comparing the reduced forms of $S_A,S_B$ given
below we find the map between the A,B types in 9 dimensions. 
\begin{equation}
\begin{array}{ll}
S_B= & \bar{\gamma}_{\bar{\alpha}\bar{\beta}}^9\,\,\left( c\bar{\tau}%
_i\right) ^{\bar{a}\bar{b}}P_9^i+\bar{\gamma}_{\bar{\alpha}\bar{\beta}}^\mu
\,\,\left( c\bar{\tau}_i\right) ^{\bar{a}\bar{b}}P_\mu ^i \\ 
& +\left( \bar{\gamma}^9\,\bar{\gamma}^{\mu _1\mu _2}\right) _{\bar{\alpha}%
\bar{\beta}}\,\,c^{\bar{a}\bar{b}}\,Y_{9\mu _1\mu _2} \\ 
& +\bar{\gamma}_{\bar{\alpha}\bar{\beta}}^{\mu _1\mu _2\mu _3}\,\,c^{\bar{a}%
\bar{b}}\,Y_{\mu _1\mu _2\mu _3} \\ 
& +\left( \bar{\gamma}^9\bar{\gamma}^{\mu _1\cdots \mu _4}\right) _{\bar{%
\alpha}\bar{\beta}}\,\,\,\left( c\bar{\tau}_i\right) ^{\bar{a}\bar{b}%
}X_{9\mu _1\cdots \mu _4}^i \\ 
& i=0^{\prime },10,A=0^{\prime },1^{\prime },2^{\prime };\quad \mu
=0,1,\cdots ,8
\end{array}
\end{equation}

\begin{equation}
\begin{array}{ll}
S_A= & \left( C\gamma ^{IJ}\right) _{\alpha \beta }\,\,\,Z_{IJ}+\left(
C\gamma ^\mu \gamma ^I\right) _{\alpha \beta }\,\,\,Z_{\mu I} \\ 
& +\left( C\gamma ^{\mu _1\mu _2}\right) _{\alpha \beta }\,\,\,Z_{\mu _1\mu
_2} \\ 
& +\left( C\gamma ^{\mu _1\mu _2\mu _3}\gamma ^{0^{\prime }}\gamma ^9\gamma
^{10}\right) _{\alpha \beta }\,\,\,Z_{\mu _1\mu _2\mu _30^{\prime }910}^{+}
\\ 
& +\left( C\gamma ^{\mu _1\cdots \mu _4}\gamma ^{IJ}\right) _{\alpha \beta
}\,\,\,Z_{\mu _1\cdots \mu _4IJ}^{+} \\ 
& I=0^{\prime },10,9,\quad \mu =0,1,\cdots ,8
\end{array}
\end{equation}
These expressions are obtained by rewriting the original expressions for $%
S_A,S_B,$ that were given in terms of the 64$\times 64$ gamma matrices $%
\Gamma $ in the appendix, and specializing the indices $M=(\mu ,I)$ or $(\mu
,i)$ respectively. There is a correspondance term by term: $%
Z_{IJ}\longleftrightarrow \varepsilon _{IJK}P_9^K,$ $Z_{\mu
I}\longleftrightarrow P_\mu ^i,$ etc.. However, the 32$\times 32$ gamma
matrices of type A that multiply these coefficients are not of the direct
product form $\bar{\tau}\otimes \bar{\gamma}$ of type B. Furthermore, the
type A,B indices $I,i$ respectively label different sets of compactified
dimensions embedded in 13D ($I=0^{\prime },10,9,$ versus $i=0^{\prime
},10,A, $ where $A$ labels the 13$^{th}$ dimension$)$. Hence the T-duality
that exists between types A and B is closely related to the map provided by
the above expressions, and it involves a{\it \ ``duality'' transformation
that corresponds to relabelling some of the 13 dimensions}.

\section{Representations, BPS states}

The superalgebra in $d$ dimensions (\ref{ab}) has two types of isometries:
spacetime-like isometries $SO(c+1,1)$ and duality isometries $K$ ($\subset
U) $ in addition to the Lorentz isometry $SO(d-1,1).$ The classification of
the various generators has been tabulated in various dimensions $\left(
d,c\right) $ elsewhere \cite{ibbeyond}. The physical states of the theory
must be classified as the representation spaces of the superalgebra.
Therefore, it is expected that the physical states form supermultiplets
consistent with these isometries and that they reveal the structures of the
hidden dimensions and dualities displayed in (\ref{subgroups}). Some work in
this direction has been reported before \cite{ib11}\cite{ibbeyond}. Here we
describe a more systematic approach and provide examples of new BPS states
that belong in larger multiplets along with previously known BPS states. The
new element is the inclusion of quantum numbers that carry Lorentz indices.

In the case of Abelian bosonic generators, as assumed in this paper, all
representations are found by analogy to representations of standard
supersymmetry. The main new ingredient is that instead of the standard
momentum operators we now have 528 abelian operators in $S$ that are
simultaneosly diagonalizable. In previous work only the Lorentz singlet
central extensions were included in addition to the standard momentum in
seeking representations, but here we include all bosonic operators. Recall
that those that carry $p$ Lorentz indices are physically relevant for the
description of boundaries of $p$-branes in ordinary spacetime (not just in
compactified spacetime). These 528 operators are at an equal footing since
they are mixed with the momenta and with each other by the $SO(c+1,1)$ and $%
K $ isometries.

The representation space in constructed as follows. A reference state is
chosen such that it is labelled by 
\begin{equation}
|S,R_c>\text{or}\,\,\,|S,R_K>  \label{reference}
\end{equation}
where $S$ represents the eigenvalues of all commuting 528 bosonic
generators, and $R_{c,K}$ is a representation of the isometry $%
SO(d-1,1)\otimes SO(c+1,1)$ or a representation of $SO(d-1,1)\otimes K$
respectively (as discussed in \cite{ibbeyond} $R_c$ must form a collection
of ireducible representations that can be expanded in terms of $R_K$ and
vice-versa). Then all possible powers of the fermionic generators are
applied to the reference state in order to obtain the full supermultiplet.

For ``long'' multiplets there are $2^{32/2}$ combinations of linearly
independent powers of fermionic operators applied on the reference state
(fermionic plus bosonic spinor representations of SO(32)). Since the
reference state has dimension $\dim (R_{c,K}),$ then the dimension of the
full supermultiplet is $2^{16}\times $ $\dim (R_{c,K}).$ Furthermore, since
each supergenerator is classified under $SO(d-1,1)\otimes SO(c+1,1)$ or $%
SO(d-1,1)\otimes K,$ it is straightforward to obtain the representation
content of each state under these groups. We will argue in the next section
that these supermultiplets hide even bigger structures associated with 12D
or 13D.

Some of the irreducible supermultiplets are shorter than the naive counting
would indicate. This happens whenever there is a linear combination of
fermionic generators that vanishes on the reference state 
\begin{equation}
\epsilon _{a(k)}^\alpha Q_\alpha ^a\,\,|S,R_{c,K}>=0,\,\,\quad k=0,1,\cdots
,n  \label{zerosuper}
\end{equation}
The supermultiplets associated with such reference states are the BPS-type
states. This gives the analog of shorter multiplets of ordinary
supersymmetry with central extensions, but here the possibilities are much
richer since there are many more central extensions. With only Lorentz
singlet central extension the shorter multiplets look like representations
of 1/2, 1/4, 1/8,$\cdots $ supersymmetry. But by including all possible
central extensions we find that it is possible to obtain shorter
supermultiplets as if the supersymmetry contains any number of generators
from zero to 32, i.e. $n$ can be anything.

It is important to emphasize that since the momentum $P_\mu $ (and in
particular the mass) is mixed with all other 528 quantum numbers under $%
SO(c+1,1)$, {\it these multiplets can contain states of different masses}.
We see then that the multiplets have plenty of information about the hidden
dimensions or duality symmetries of the theory. The more familiar string
states at various excitation levels are part of the multiplet; the
additional states needed to complete the multiplet become the prediction of
S-theory. In previous work some simple examples in this direction were
provided \cite{ib11}\cite{ibbeyond}.

Now we give the covariant criteria, consistent with all the isometries, for
the presence BPS-like supermultiplets . Since (\ref{zerosuper}) must hold,
then it implies that the 32$\times 32$ matrix $S$ must have zero eigenvalues
with multiplicity $n$%
\begin{equation}
\epsilon _{a(k)}^\alpha \left\{ Q_\alpha ^a,Q_\beta ^b\right\}
\,\,|S,R_{c,K}>=0\quad \rightarrow \quad \epsilon _{a(k)}^\alpha S_{\alpha
\beta }^{ab}=0
\end{equation}
Therefore, the determinant vanishes 
\begin{equation}
\det \left( S\right) =0.
\end{equation}
By writing out the secular equation $\det \left( S-\lambda \right) =0$ the
multiplicity of the zero eigenvalue (i.e. $n)$ can be determined. In our
notation the energy term in $S\,\,\,$is proportional to the identity $S\sim $
$P_0+\cdots $ as described at the end of section-3.2. Hence adding the $%
\lambda $ is superfluous; we can instead count the multiplicity of the
energy eigenvalue at which $\det \left( S\right) =0.$ This condition is
consistent with all the isometries, and therefore the collection of all
BPS-like states that satisfy it must form a shorter supermultiplet of the
superalgebra and of the isometries. It is worth emphasizing that all the
information about the multiplet is contained in the reference state (\ref
{reference}).

\section{Higher dimensions and BPS states}

As discussed in section-3, the form of $S$ in $\left( d,c\right) $
dimensions is a rewriting of the original $S_{A,B}$ embedded in 12 or 13
dimensions. Furthermore, the criteria for the BPS-like states (as well as
for long multiplets) are consistent with the higher symmetries that are
displayed by the original $S_{A,B}.$ Therefore, the long or shorter
supermultiplets that are identified in any dimension must also be consistent
with the symmetry structure of the hidden 12 or 13 dimensions, i.e. SO(10,2)$%
_A$ or SO(9,1)$\otimes $SO(2,1)$_B$. Some examples are provided below.

\subsection{From Type IIA superstring to 12D supergravity}

As is well known by now the black holes of type IIA superstring can be
thought of as Kaluza-Klein states of 11D supergravity compactified to 10D.
In our language these BPS states correspond to a reference state 
\begin{equation}
|p_\mu ,p_{10};\,\,R_{c,K}={\bf 1>}
\end{equation}
where in addition to momentum, the only non-zero central extension in 10D is
the quantized 11th momentum $p_{10}=n/R$ where $R$ is the radius of
compactification (related to the coupling constant as argued by Witten \cite
{witten}). The remaining bosonic 517 $\left( =528-11\right) $ central
extensions are set equal to zero. Then (\ref{type2},\ref{type2a}) and the
BPS conditions simplify to 
\begin{equation}
\begin{array}{l}
S_A=p_\mu \left( C\gamma ^\mu \right) +p_{10}\left( C\gamma ^{10}\right)  \\ 
\det \left( S_A\right) \sim \left( p_0^2-\vec{p}^2-p_{10}^2\right)
^{16}=\left( M_{10}-p_{10}\right) ^{16}\left( M_{10}+p_{10}\right) ^{16} \\ 
\det S_A=0\quad \longleftrightarrow \quad M_{10}=\left| p_{10}\right| ,\quad 
\text{multiplicity}=16
\end{array}
\label{11d}
\end{equation}
where $M_{10}$ is the mass in 10D, $M_{10}^2=p_0^2-\vec{p}^2.$ 16
supersymmetries vanish and 32-16=16 act non-trivially. Therefore, the
dimension of this short supermultiplet is $2^{16/2}=256,$ consisting of 2$^7$
bosons plus 2$^7$ fermions. This has the same content as the degrees of
freedom of 11D supergravity. As is well known, the presence of the 11th
momentum $p_{10},$ as a central extension, indicates the presence of the
11th dimension. In addition, 11D manifests itself in the 256 dimensional
supermultiplet of states, since this multiplet is constructed from the
direct products of the 32 supercharges that form a spinor basis for 11D.

Now we discuss the hidden 12D structure. The first hint is that the 32
supercharges form a basis for the chiral spinor representation for 12D.
Therefore, the multiplicity 256 is consistent with 12D Lorentz
transformations SO(10,2). However, there is no 12th momentum. Indeed, as
discussed in section-3, a 12th momentum should not be expected, since it
does not appear in the superalgebra of type A (\ref{type2},\ref{type2a}).
Instead one should seek the central extensions $Z_{M_1M_2},\,$ $Z_{M_1\cdots
M_6}^{+}.$ Consider a reference state labelled by 
\begin{equation}
|Z_{M_1M_2}\neq 0,\,Z_{M_1\cdots M_6}^{+}=0\,\,;\,\,R_{c,K}={\bf 1>}
\end{equation}
with the further special condition $\left( Z^3\right) _{M_1M_2}=0$ which is
still SO(10,2) covariant (contractions of indices requires the (10,2)
metric). A form of $Z_{M_1M_2}$ that satisfies this requirement is 
\begin{equation}
\begin{array}{l}
Z_{M_1M_2}=\Lambda _{M_1}^{\,\,\,\,\,N_1}Z_{N_1N_2}^0\Lambda
_{\,\,\,\,M_2}^{N_2} \\ 
\,Z_{M_1M_2}^0=\left( 
\begin{array}{lll}
0 & p_{10} & p_\mu  \\ 
-p_{10} & 0 & 0 \\ 
-p_\mu  & 0 & 0
\end{array}
\right) 
\end{array}
\,\,\,\,
\end{equation}
where $\Lambda $ is a SO(10,2) boost and $Z_{M_1M_2}^0$ is the solution in (%
\ref{11d}). This form is equivalent to the cross product of two 12D vectors
that are orthogonal and one of them is null in 12D 
\begin{equation}
\begin{array}{l}
Z_{M_1M_2}=\frac 12\left( \tilde{P}_{M_1}\tilde{P}_{M_2}^{\prime }-\tilde{P}%
_{M_2}\tilde{P}_{M_1}^{\prime }\right) , \\ 
\tilde{P}\cdot \tilde{P}^{\prime }=\tilde{P}\cdot \tilde{P}=0.
\end{array}
\end{equation}
The tildas are used to emphasize that these are 12D vectors. $S_A$ and its
determinant have the form 
\begin{equation}
\begin{array}{l}
S_A=C\gamma ^{M_1M_2}\tilde{P}_{M_1}\tilde{P}_{M_2}^{\prime } \\ 
\det S_A=\left( \tilde{P}^2\tilde{P}^{\prime 2}-(\tilde{P}\cdot \tilde{P}%
^{\prime })^2\right) ^{16}
\end{array}
\label{crossform}
\end{equation}
Therefore, the zero eigenvalue is 16-fold degenerate. Written in this form
the reference state is covariant under SO(10,2), but yet it is equivalent to
the 11D reference state in (\ref{11d}) up to a SO(10,2) boost. Combining
this reference state with the fact that the 32 supercharges form a spinor
representation of SO(10,2), the resulting supermultiplet with a 256
degeneracy (just as in 11D supergravity) must also be consistent with
SO(10,2).

Therefore I conjecture that there should be a reformulation (or
generalization perhaps by including auxiliary fields) of 11D supergravity
that is consistent with more hidden dimensions, is SO(10,2) covariant and
contains the same 256 physical components of fields that are present in 11D
supergravity. However, for covariance, the fields should be allowed to
depend on more than 11 dimensions, and be consistent with the covariant
central extensions given above in the form of equations of motion for the
fields 
\begin{equation}
\tilde{P}^2\,\phi +\cdots =0,\quad \tilde{P}\cdot \tilde{P}^{\prime
}\,\,\phi +\cdots =0.
\end{equation}
Such a reformulation would provide one of the simplest low energy effective
field theories that describe a sector of the fundamental theory consistently
with SO(10,2). In view of the present remarks it may be useful to revive
some old attemps in such a direction \cite{berg}. Toward this goal perhaps a
first step should be generalizing the superparticle action with additional
degrees of freedom in 12D, such as $\tilde{P},\tilde{P}^{\prime }$ and
superpartners. The canonical quantization of of such a generalized
superparticle should yield the specialized form of the superalgebra in (\ref
{type2a}) 
\begin{equation}
\left\{ Q_\alpha ,Q_\beta \right\} =\left( C\gamma ^{M_1M_2}\right) _{\alpha
\beta }\tilde{P}_{M_1}\tilde{P}_{M_2}^{\prime }.
\end{equation}

The form (\ref{crossform}) can satisfy $\det S_A=0$ with a slightly less
constrained SO(10,2) vectors $\tilde{P},\tilde{P}^{\prime }$ . This
corresponds to a larger class of solutions that are not connected to (\ref
{11d}) by SO(10,2) boosts.

\subsubsection{More 11D$\longleftrightarrow $12D solutions}

It is possible to display some special solutions with more central
extensions consistent with 11D, and covariantized to 12D by boosts. For
example, using 11D gamma matrices as in (\ref{type2a}) one may take 
\begin{equation}
S_A=C\not{P}+C\not{P}^{\prime }\not{P}+C\not{X}_1\not{X}_2\not{X}_3\not{X}_4%
\not{P}
\end{equation}
where $\not{P}\equiv P\cdot \gamma ,\,\not{P}^{\prime }\equiv P^{\prime
}\cdot \gamma ,\,\not{X}_i\equiv X_i\cdot \gamma $ are 11D vectors dotted
with gamma matrices. To insure the antisymmetry of $Z_{\mu \nu },X_{\mu
_1\cdots \mu _5},$ the vectors are taken othogonal to each other.
Furthermore, taking $P_\mu $ lightlike in 11D (i.e. $M_{10}=\left|
p_{10}\right| $ as in (\ref{11d}) ) guarantees that the BPS condition is
satisfied 
\begin{equation}
\det S_A=\det \left( C+C\not{Z}+C\not{X}_1\not{X}_2\not{X}_3\not{X}_4\right)
\,\,\det \left( \not{P}\right) =0.
\end{equation}
Since 
\begin{equation}
\det \left( \not{P}\right) =\left( M_{10}+p_{10}\right) ^{16}\left(
M_{10}+p_{10}\right) ^{16},
\end{equation}
the multiplicity of the zero eigenvalue is again 16. In this case the
reference state has more non-zero central extensions describing more
complicated p-branes. These are probably related to each other by various
dualities.

Any 11D solution can be boosted to a 12D covariant form by applying an
overall SO(10,2) transformation and then identifying the $Z_{M_1M_2},\,\,$ $%
Z_{M_1\cdots M_6}^{+}.$ In the present solution both of these are non-zero,
albeit of special forms rather than being the most general. It is because of
their special form that the degeneracy of the zero eigenvalue is still 16.
With more general forms the degeneracy (and hence the size of the shorter
supermultiplets) would be different.

\subsubsection{Excited levels}

The excited states of type IIA (perturbative) string give only a subset of
the states of the full secret theory. We have conjectured in the past that
the correct set should correspond to supermultiplets in 11D and found some
evidence for this \cite{ib11}\cite{ibbeyond}. We now modify this conjecture
because we expect the full theory to be consistent with representations of
the superalgebra as described in section-3. Then the excited levels should
be classified according to reference states with non-trivial $R_{c,K},$ but
the same form of $S_A$ as the base$.$ The well known (perturbative) excited
string states should fill part of these multiplets. The remainder of the
multiplet is a prediction about the properties of the underlying secret
theory, as seen in examples in our previous work.

\subsection{Example of a vector central extension}

Consider superstring theory of type II compactified to 9D. The base state is
labelled with a 9D momentum $p_\mu ,$ a quantized Kaluza-Klein momentum $k_9,
$ and a winding number $w_9.$ If these are the only non-trivial central
extensions then they are embedded in the 12D $Z_{MN}$ as follows (where the
order of the indices is taken as $M,N=9,10,0^{\prime },0,1,2,\cdots ,8)$ 
\begin{equation}
Z_{MN}=\left( 
\begin{array}{lllllll}
0 & -w_9 & -k_9 & 0 & 0 & \cdots  & 0 \\ 
w_9 & 0 & 0 & 0 & 0 & \cdots  & 0 \\ 
k_9 & 0 & 0 & p_0 & p_1 & \cdots  & p_8 \\ 
0 & 0 & -p_0 & 0 & 0 & \cdots  & 0 \\ 
0 & 0 & -p_1 & 0 & 0 & \cdots  & 0 \\ 
\vdots  & \vdots  & \vdots  & \vdots  & \vdots  & \ddots  & 0 \\ 
0 & 0 & -p_8 & 0 & 0 & 0 & 0
\end{array}
\right) 
\end{equation}
while $Z_{M_1\cdots M_2}^{+}=0.$ This form is covariant under Lorentz
transformations SO(8,1) but non-covariant under the isometries of the
superalgebra given in (\ref{subgroups}). The perturbative string states are
well known at all excited levels. The non-perturbative black hole states
satisfy the well known BPS condition $M_9=\left| k_9\pm w_9\right| ,$ where $%
M_9^2=p_0^2-\vec{p}^2.$ This corresponds to requiring the vanishing of the
determinant 
\begin{equation}
\det S=(M_9-w_9-k_9)^8(M_9-w_9+k_9)^8(M_9+w_9+k_9)^8(M_9+w_9-k_9)^8
\end{equation}
which has an 8-fold degeneracy for the zero eigenvalue. This means that 8
supergenerators vanish and 32-8=24 of them act non-trivially on the
reference state, giving a well known shorter supermultiplet of dimension $%
2^{24/2}=2_{bosons}^{11}+2_{fermions}^{11}.$

First I generalize this by including the central extension $P_{10}$ which is
a Lorentz singlet, and whose presence is required by U-duality \footnote{%
In this case $U=SL(2)\times SO(1,1)$ which has the maximal compact subgroup $%
K=SO(2)\times Z_2$. Under $K$ the Lorentz singlet central extensions form a
doublet ($k_9,k_{10})$ plus a singlet $w_9.$}$.\,\,$Then 
\begin{equation}
Z_{MN}=\left( 
\begin{array}{lllllll}
0 & -w_9 & -k_9 & 0 & 0 & \cdots  & 0 \\ 
w_9 & 0 & -k_{10} & 0 & 0 & \cdots  & 0 \\ 
k_9 & k_{10} & 0 & p_0 & p_1 & \cdots  & p_8 \\ 
0 & 0 & -p_0 & 0 & 0 & \cdots  & 0 \\ 
0 & 0 & -p_1 & 0 & 0 & \cdots  & 0 \\ 
\vdots  & \vdots  & \vdots  & \vdots  & \vdots  & \ddots  & 0 \\ 
0 & 0 & -p_8 & 0 & 0 & 0 & 0
\end{array}
\right) 
\end{equation}
gives 
\begin{equation}
\det S=\left( \left( M_9-w_9\right) ^2-k_9^2-k_{10}^2\right) ^8\left( \left(
M_9+w_9\right) ^2-k_9^2-k_{10}^2\right) ^8
\end{equation}
which has manifest $K\otimes SO(8,1)$ symmetry. The degeneracy of the zero
eigenvalue is still 8, hence the size of the supermultiplet is $%
2_{bosons}^{11}+2_{fermions}^{11}$ but the base has one more quantum number,
and it displays the explicit $K$ isometry. The mass formula is  $K$
invariant.
\begin{equation}
M_9=\left| w_9\pm \sqrt{k_9^2+k_{10}^2}\right| 
\end{equation}

So far this is insufficient to also display the $SO(c+1,1)=SO(2,1)$ isometry
given in (\ref{subgroups}). As explained there, the $SO(2)$ subgroup is the
same as the one appearing in $K.$ The $SO(2,1)$ transformations mix the
indices $M=0^{\prime },9,10.$ When these are applied to the $Z_{MN}$ above
they require the more general covariant form \footnote{$Z_{M_1\cdots
M_2}^{+}=0$ is still consistent with the isometries so far, but without
turning on this central extension as well, the full 12D covariance remains
hidden.}

\begin{equation}
Z_{MN}=\left( 
\begin{array}{lllllll}
0 & -w_9 & -k_9 & z_{90} & z_{91} & \cdots  & z_{98} \\ 
w_9 & 0 & -k_{10} & z_{100} & z_{101} & \cdots  & z_{108} \\ 
k_9 & k_{10} & 0 & p_0 & p_1 & \cdots  & p_8 \\ 
-z_{90} & -z_{100} & -p_0 & 0 & 0 & \cdots  & 0 \\ 
-z_{91} & -z_{101} & -p_1 & 0 & 0 & \cdots  & 0 \\ 
\vdots  & \vdots  & \vdots  & \vdots  & \vdots  & \ddots  & 0 \\ 
-z_{98} & -z_{108} & -p_8 & 0 & 0 & 0 & 0
\end{array}
\right) 
\end{equation}
The additional $z_{9\mu },z_{10\mu }$ are Lorentz vectors that are
interpreted as positions of end points of strings in 9D as explained in
section-2. Therefore, they must be space-like vectors. Together with the
timelike $p_\mu \equiv z_{0^{\prime }\mu }$ they form a triplet $z_{i\mu }$
of $SO(2,1)$ $.$ When these are included in the reference state we obtain a
supermultiplet consistent with $SO(2,1)\otimes SO(8,1).$ The determinant of $%
S\sim C\gamma ^{MN}Z_{MN}\,$\thinspace \thinspace is 
\begin{equation}
\det S=\left( TrZ^4-\left( Tr\frac{Z^2}2\right) ^2\right) ^8
\end{equation}
The BPS determinant condition is evidently invariant under $SO(2,1)\otimes
SO(8,1),$ the degeneracy of the zero eigenvalue is 8 and the size of the
supermultiplet is $2_{bosons}^{11}+2_{fermions}^{11}$. This supermultiplet
is consistent with the isometries $SO(2,1)\otimes SO(8,1)$ (as well as with $%
K)$ since all supercharges and all central extensions are complete
multiplets of the isometries (for their classification in every dimensions
see \cite{ibbeyond}).

In this way we have extended the perturbative superstring multiplets to the
larger supermultiplets containing non-perturbative states of the secret
theory consistent with 12D or 13D.

\section{Conclusions}

In the examples of section-4 only some of the central extensions were turned
on. This was sufficient to illustrate the method of investigation as well as
the relevance and properties of the Lorentz non-singlet central extensions.
Even with this limited set of examples it is clear that the spectrum and
multiplet structure of the secret theory is much richer than previously
thought. More work is needed to find all the solutions of $detS_{A,B}=0$ and
identify all the distinct shorter multiplets. The number of supersymmetries
on BPS states in the examples of this paper were 8,16, but all numbers from
0 to 32 are possible in more general examples. The algebraic approach
outlined in this paper seems to be sufficiently powerful to elucidate some
of the non-perturbative global properties of the secret theory. For example,
by starting with known perturbative string states at any excitation level
and using the multiplets of S-theory one can in principle make predictions
on the spectrum of the secret theory. Some preliminary examples of this type
were given before \cite{ib11}\cite{ibbeyond}. Similar considerations should
also apply to scattering amplitudes, etc. In S-theory one finds that there
are up to 13 hidden dimensions, some of which remain hidden from the point
of view of perturbative approaches involving $p$-branes. One of the
appealing aspects of the S-theory approach is to treat all 528 bosonic
generators on an equal footing, thus elucidating the duality and hidden
dimensions as simple consequences of the isometries of the maximally
extended superalgebra.

There should be many variations of S-theory by taking some of the 528
bosonic generators to be non-abelian. Some of them may also have non-trivial
commutation relations with the supergenerators\cite{sezgin}. Such variations
of the superalgebra must be related to the geometrical properties of the
background in which the p-branes propagate, as opposed to the flat
background assumed in the present paper. Evidently the representation theory
of the corresponding superalgebra will be more difficult but more
interesting. In this way it should be possible to find relations between the
results of M-,F- and S-theories.

\section{Appendix}

The gamma matrices in 12D with signature (10,2), or in 13D with signature
(11,2), may be given explicitly in the following 64$\times 64$ purely real
(Majorana) representation, using direct products of Pauli matrices 
\begin{equation}
\begin{array}{lll}
\Gamma _{0^{\prime }}=i\tau _2\otimes 1\otimes 1\otimes 1_8 & \quad \quad  & 
\Gamma _9=\tau _1\otimes \sigma _1\otimes \tau _3\otimes 1_8 \\ 
\Gamma _0=\tau _1\otimes i\sigma _2\otimes 1\otimes 1_8 &  & \Gamma _8=\tau
_1\otimes \sigma _1\otimes \tau _1\otimes 1_8 \\ 
\Gamma _{10}=\tau _1\otimes \sigma _3\otimes 1\otimes 1_8 &  & \Gamma
_i=\tau _1\otimes \sigma _1\otimes \tau _2\otimes g_i \\ 
\Gamma _A=\tau _3\otimes 1\otimes 1\otimes 1_8 &  & C=1\otimes i\sigma
_2\otimes 1\otimes 1_8 \\ 
\Gamma _B=1\otimes \sigma _3\otimes 1\otimes 1_8 &  & 
\end{array}
\label{gammas}
\end{equation}
where the $g_i$ are purely imaginary 8$\times 8$ antisymmetric gamma
matrices for the remaining 7 dimensions. $C$ is the charge conjugation
matrix, it has the property that $C\Gamma _M$ is symmetric for the 12D gamma
matrices, $M=0^{\prime },0,1,\cdots ,10$, or $C\Gamma _MC^{-1}=-\left(
\Gamma _M\right) ^T.$

$\Gamma _A$ is a 13$^{th}$ gamma matrix that is the product of the 12D $%
\Gamma _M$ and it anticommutes with them (i.e. analog of $\gamma _5$ in 4D). 
$\Gamma _A$ commutes with $C$ and $\Gamma _B$ . The chiral projector $\frac
12\left( 1+\Gamma _A\right) $ serves to project to the 32$\times 32$
subspace that is of interest for the type IIA sector of the theory. Since
the chirally projected sector distinguishes the 13$^{th}$ gamma matrix the
maximal covariance is broken down from 13D to 12D in the type A sector.

When the antisymmetric products of $p$ gamma matrices $\Gamma _{M_1M_2\cdots
M_p}$ are multiplied by the projector $1+\Gamma _A$ only the 12D covariant 66%
$\rightarrow $ $\frac 12\left( 1+\Gamma _A\right) C\Gamma _{M_1M_2\text{ }}$%
and 462 $\rightarrow \frac 12\left( 1+\Gamma _A\right) C\Gamma
_{M_1M_2\cdots M_6}$ are symmetric matrices. Therefore only the 2-index and
self dual (in 12D) six-index tensors can appear in the 12D type A
superalgebra. Thus $S_A$ is a linear combination of these as in (\ref{type2}%
). In this 32$\times 32$ subspace we may replace each $\Gamma _M$ by $\frac
12\left( 1+\Gamma _A\right) \Gamma _M\frac 12\left( 1-\Gamma _A\right) $ $%
\rightarrow $ $\gamma _M,$ where we denote the 32$\times 32$ gamma matrices $%
\gamma _M$ by $\gamma _M=(1,\gamma _m),$ with $\gamma _{0^{\prime }}=1$ and $%
\gamma _m,\,\,m=0,1,\cdots ,10$ given by omiting the first $\tau _i$ factors
in the expressions of the other 11 $\Gamma _m$ given above, i.e. $\gamma
_0=i\sigma _2\otimes 1\otimes 1_8=C,\;$ $\gamma _{10}=\sigma _3\otimes
1\otimes 1_8$, etc.. This form may be used in (\ref{type2a}) to simplify it
to the 11D notation of (\ref{type2}).

$\frac 12\left( 1+\Gamma _B\right) $ is the projector to the 32 dimensional
subspace relevant for the type-B sector of the theory. $\Gamma _B$ is the
product of the usual 10D $\Gamma _\mu $ gamma matrices $\Gamma _B\sim \Gamma
_0\Gamma _1\cdots \Gamma _9$. One can also write 
\begin{equation}
\Gamma _B\sim \Gamma _{0^{\prime }}\Gamma _{10}\Gamma _A\sim \Gamma _0\Gamma
_1\cdots \Gamma _9.
\end{equation}
$\Gamma _B$ commutes with each $\Gamma _i\equiv $ $\left( \Gamma _{0^{\prime
}},\Gamma _{10},\Gamma _A\right) $ and anticommutes with $C$ and each $%
\Gamma _\mu =\left( \Gamma _0,\Gamma _1,\cdots ,\Gamma _9\right) $%
\begin{equation}
\ \left[ \Gamma _B,\Gamma _i\right] =0,\quad \left\{ \ \Gamma _B,\Gamma _\mu
\right\} =\left\{ \ \Gamma _B,C\right\} =0.
\end{equation}
Therefore this projection breaks the symmetry from 13D to 10D$\otimes $3D
since it treats the 3D differently than the 10D. The three $\Gamma _i$ may
be regarded as the gamma matrices for a 3D hidden Minkowski space just as
the ten $\Gamma _\mu $ are the gamma matrices for the 10D Minkowski space.
This extra space is evidently related to the geometrical origin of the
SL(2,R) symmetry of the type-B sector.

$S_B$ is constructed from 528 linearly independent symmetric 32$\times 32$
matrices of type-B. In 64$\times 64$ notation these are given by 
\begin{equation}
\begin{array}{lll}
\frac 12\left( 1+\Gamma _B\right) C\Gamma _{\mu _1\mu _2\mu _{31}\mu _4\mu
_5}\left( \Gamma _A\Gamma _i\right) \quad  & :\quad  & \frac 12\frac{%
10\times 9\times 8\times 7\times 6}{1\times 2\times 3\times 4\times 5}\times
3=378 \\ 
\frac 12\left( 1+\Gamma _B\right) C\Gamma _{\mu _1\mu _2\mu _{31}}\left(
\Gamma _A\right)  & : & \frac{10\times 9\times 8}{1\times 2\times 3}=120 \\ 
\frac 12\left( 1+\Gamma _B\right) C\Gamma _{\mu _1}\left( \Gamma _A\Gamma
_i\right)  & : & 10\times 3=30
\end{array}
\label{typebgammas}
\end{equation}
where the chirally projected five index gamma matrices are self dual in 10D.
In the chirally projected B sector these matrices reduce to 32$\times 32$
blocks which may be conveniently written in the form of direct products of 2$%
\times 2$ times 16$\times 16$\thinspace matrices $\bar{\tau}\otimes \bar{%
\gamma}$ as in (\ref{type2b}), where the 2$\times 2$ part comes directly
from the first factor and the 16$\times 16$ part comes from the last three
factors in the gamma matrix expressions $\Gamma $ in (\ref{gammas},\ref
{typebgammas}).

One may consider SO(1,2)=SL(2,R) rotatios of $S_B$ in the extra 3D subspace,
leaving unaffected the usual 10 dimensions. To make the connection to 13D we
give it in the form of rotations in the 64$\times 64$ spinor space 
\begin{equation}
\delta S_B=\left[ \epsilon ^{ij}\Gamma _{ij},S_B\right] .
\end{equation}
One can show that 
\begin{equation}
\epsilon ^{ij}\Gamma _{ij}\left( \frac 12\left( 1+\Gamma _B\right) C\Gamma
_{\mu _1\cdots \mu _p}\right) =\left( \frac 12\left( 1+\Gamma _B\right)
C\Gamma _{\mu _1\cdots \mu _p}\right) \Gamma _A\epsilon ^{ij}\Gamma
_{ij}\Gamma _A
\end{equation}
when $p=odd.$ Using this identity we see easily that the following
commutators simplify 
\begin{eqnarray}
&&\left[ \epsilon ^{ij}\Gamma _{ij},\left( \frac 12\left( 1+\Gamma _B\right)
C\Gamma _{\mu _1\cdots \mu _p}\left( \Gamma _A\Gamma _{k\cdots }\right)
\right) \right]   \nonumber \\
&=&\frac 12\left( 1+\Gamma _B\right) C\Gamma _{\mu _1\cdots \mu _p}\Gamma
_A\left[ \epsilon ^{ij}\Gamma _{ij},\left( \Gamma _{k\cdots }\right) \right] 
\end{eqnarray}
The last commutator is just the rule for performing rotations in the 3D
subspace. This shows that only the 3D indices rotate under these SO(1,2)
rotations embedded in 13D rotations, and hence verifies that the
construction of the 528 matrices (\ref{typebgammas}) is the right one.
Furthermore, this result is consistent with using the direct product
notation of (\ref{type2b}). The construction of (\ref{typebgammas}) is
useful because it exhibits the precise embedding of the type-B space in the
spinor space of 13D.

From the constructions given above we see that the type A and type B are
differents projections within the same 64$\times 64$ spinor space of 13D.
Hence the duality of the type-A and type-B sectors of the theory has its
origins in the spinor space for 13D .

\end{document}